\documentclass[12pt,preprint]{aastex}

\shorttitle{Herbig Ae/Be Candidate Star Photometry}
\shortauthors{Doering \& Meixner}

\begin{document}

\title{Near-Infrared Photometric Survey of Herbig Ae/Be Candidate Stars}

\author{Ryan L. Doering\altaffilmark{1}
\& Margaret Meixner\altaffilmark{2}}

\altaffiltext{1}{Department of Astronomy, University of Wisconsin at Madison, 475 North Charter Street, Madison, WI 53706, USA}
\altaffiltext{2}{Space Telescope Science Institute, 3700 San Martin Drive, Baltimore, MD 21218, USA}

\email{doering@astro.wisc.edu}

\begin{abstract}

We report near-infrared photometric measurements of 35 Herbig Ae/Be candidate stars obtained with direct imaging and aperture photometry.  Observations were made through the broadband $J$, $H$, and $K$$^{\prime}$ filters, with each source imaged in at least one of the wavebands.  We achieved subarcsecond angular resolution for all observations, providing us with the opportunity to search for close binary candidates and extended structure.  The imaging revealed five newly identified binary candidates and one previously resolved T Tauri binary among the target sources with separations of $\la$2$\farcs$5.  Separate photometry is provided for each of the binary candidate stars.  We detect one extended source that has been identified as a protoplanetary nebula.  Comparing our magnitudes to past measurements yields significant differences for some sources, possibly indicating photometric variability.  $H$-band finding charts for all of our sources are provided to aid follow-up high-resolution imaging.

\end{abstract}


\section{Introduction}

Herbig Ae/Be stars are thought to be in a pre-main sequence (PMS) evolutionary stage, and are considered the intermediate-mass ($\sim$2--10 $M_{\sun}$) analogs of the T Tauri stars.  They were originally classified by \citet{herbig60} as emission line stars of spectral type A or earlier that lie in an obscured region, and illuminate a reflection nebula in their vicinity.  More recent studies have broadened the selection criteria in order to include stars that are most likely of a similar type \citep[e.g.,][]{fink84, the94, malfait98}.  

The detection of an infrared (IR) excess indicating the presence of circumstellar dust is generally accepted as a key criterion for membership among the Herbig Ae/Be stars.  This IR excess is usually seen as a second peak in a star's spectral energy distribution (SED).  Detailed models of a star's SED are used to determine circumstellar dust properties such as mass, geometry, and grain size and composition \citep[e.g.,][]{hillenbrand92, malfait98}.  Therefore, obtaining photometric measurements for the purpose of creating and analyzing SEDs of Herbig Ae/Be candidate stars is necessary in order to better understand these sources.

The near-IR spectral regime provides us with critically important information about the characteristics of Herbig Ae/Be star circumstellar environments.  For younger Herbig Ae/Be stars, dust may reside at or near the sublimation radius where it is much warmer than the dust located further from the star, creating a near-IR excess.  Moreover, the heating from direct stellar radiation causes the inner rim to puff up giving rise to the near-IR bump seen in some Herbig Ae/Be star SEDs, and resulting in partially or fully shadowed circumstellar disks \citep{natta01,dullemond01}.  Since direct imaging has not yet resolved the inner rim of any Herbig Ae/Be sources, photometric measurements remain one of the best options for studying this geometric effect.  In addition to the usual dust grain radiative heating, small grains may be transiently heated by stellar ultraviolet (UV) flux and contribute to near-IR excess emission.  The lack of a near-IR excess may indicate an older, more evolved Herbig Ae/Be star in which an inner clearing has opened up in the dust distribution.

Near-IR photometric variability has been observed for a number of Herbig Ae/Be stars \citep[e.g.,][]{davies90, eiroa02}.  The variability appears irregular and with ranges on the order of a few to several tenths of a magnitude in the $J$, $H$, and $K$ bands.  Suggested mechanisms generally fall into two categories: photospheric activity and variations in the circumstellar environment.  The latter category includes both dust occultations of the central star and variable accretion, and implies clumpy dust distributions.  Also, if the dust is confined to a circumstellar disk, the source must be viewed nearly edge-on from the Earth for the dust to pass through our line of sight toward the star, and therefore, variability is a function of geometry.  Distinguishing between these mechanisms in individual sources has proved difficult, and the observed variability may be due to combinations of them.

A number of previous high-resolution studies have found significant binary frequencies (25\%--68\%) for samples of Herbig Ae/Be stars \citep[e.g.,][]{leinert97, baines06}.  These multiple systems provide information about the star formation history of Herbig Ae/Be stars, and allow us to study how binarity affects the circumstellar matter around these presumably coeval stars, including disk-disk interactions.  Near-IR imaging can be a useful tool for identifying cool companions around brighter (B and A) stars as well since the peak flux of stars of later spectral types falls in this spectral range.  Therefore, photometric imaging studies capable of producing separate SEDs for each binary component are crucial to better understand these systems.

\citet{vieira03} have compiled a catalog of 131 Herbig Ae/Be stars, including 108 newly identified sources from the Pico dos Dias Survey (PDS).  They provide optical photometry, spectral types, H$\alpha$ and forbidden line spectroscopy, and distance estimates.  By combining these measurements with Two Micron All Sky Survey (2MASS) $JHK$ photometry and {\it IRAS} far-IR fluxes, a robust data set becomes available for these sources.  However, the optical photometry was collected with a photometer, and 2MASS uses a 2$\arcsec$ pixel size, meaning high-resolution images do not exist for most of the stars in this catalog.

We have imaged 35 Herbig Ae/Be candidate stars from the \citet{vieira03} catalog in the near-IR in order to search for multiple systems, resolve extended structure, and collect photometry for use in SED construction by improving on previous studies such as 2MASS with higher angular resolution.  The selected sources are accessible from the northern hemisphere and are approximately representative of the complete PDS Herbig Ae/Be candidate catalog in terms of spectral type and H$\alpha$ line profile type distribution, and the presence of forbidden lines.  The stars of spectral type O, B, A, and F in our source list make up 3\%, 46\%, 43\%, and 8\% of our sample, respectively.  The PDS sample breaks down by spectral type as 5\% O, 48\% B, 37\% A, and 10\% F.  \citet{vieira03} found a correlation between H$\alpha$ line profile type and the contribution to a source's SED from circumstellar matter.  Sources with Type I profiles typically have a relatively small IR excess, Type II profile sources have substantial circumstellar contributions to their SEDs, and Type III and Type IV profile sources have intermediate contributions.  Our sample contains 29\% Type I, 40\% Type II, and 31\% Type III and IV sources compared to 24\% Type I, 43\% Type II, and 33\% Type III and IV sources in the PDS sample.  The authors also report that stars with [O I] and/or [S II] forbidden lines in their spectra tend to have significant circumstellar contributions to their SEDs.  The presence of forbidden lines is seen in the spectra of 46\% of the sources in the PDS sample.  Our forbidden line frequency is 34\%, and therefore deviates from the complete sample more than the other characteristics.  As noted by \citet{vieira03}, these values are likely lower limits due to the lack of high-resolution spectra for some sources.  Finally, a few of the brightest objects were excluded from our study due to source saturation.  In Section 2, we describe the observations, data reduction procedure, and the photometric measurements.  A discussion of the results with notes on the individual sources is given in Section 3.  The main conclusions are summarized in Section 4.

\section{Observations, Data Reduction, and Photometry}

\subsection{Observations}

Observations were carried out at the 3.5 m WIYN telescope during three runs in 2003 December, 2004 November, and 2005 September using the Near-Infrared Imager (NIRIM; Meixner et al. 1999).  NIRIM's detector is a Rockwell NICMOS3 Hg:Cd:Te 256 $\times$ 256 pixel array.  We chose the smallest available plate scale of 0$\farcs$186 pixel$^{-1}$, giving a 47$\farcs$6 $\times$ 47$\farcs$6 field of view, in order to achieve the highest spatial resolution possible in our images.  We observed through NIRIM's broadband $J$ ($\lambda$ $=$ 1.257 $\mu$m, $\Delta\lambda$ $=$ 0.293 $\mu$m), $H$ ($\lambda$ $=$ 1.649 $\mu$m, $\Delta\lambda$ $=$ 0.313 $\mu$m), and $K$$^{\prime}$ ($\lambda$ $=$ 2.12 $\mu$m, $\Delta\lambda$ $=$ 0.348 $\mu$m) filters.

A log of our observations listing the nightly average seeing values and weather conditions is shown in Table 1.  The dates correspond to the beginning of the night, local time.  The 2003 December and 2004 November dates were scheduled as full nights, while the 2005 September dates were scheduled as partial nights.  The observing runs yielded four photometric nights, three nights during which non-photometric imaging was carried out, and one night completely lost due to clouds.   Near-IR data on 35 sources were collected through at least one bandpass under photometric conditions.  Photometric standard stars were observed throughout each night in order to facilitate the magnitude calibration.  The standard star properties are listed in Table 2.  On non-photometric nights (hazy and scattered clouds), the observing strategy was to image mainly in the $H$-band to search for extended structure and binary candidates.  The observations were carried out in $H$ instead of $K^{\prime}$ in order to optimize our search for the detection of extended scattered light.  Additionally, NIRIM is most sensitive in the $H$-band due to the combination of lower sky background (relative to $K^{\prime}$) and the detector sensitivity being highest in this band.  We note that by choosing to observe in $H$ rather than $K^{\prime}$, extended thermal emission from our sources may have gone undetected.  The seeing values were determined from the average $H$-band full width at half-maximum (FWHM) of the observed standard stars during telescope focus checks throughout the night.

Several of our target sources were sufficiently bright that correctly identifying them in the field was trivial.  However, fainter sources in crowded fields were often difficult to identify without additional spatial information.  Therefore, our target source identifications were verified against optical finding charts of the fields obtained from the Digitized Sky Survey (DSS).  The telescope pointing was found to be excellent during all observing runs.  Our target sources and standard stars were observed in nine-point dither patterns on the sky using 9$\arcsec$ offsets in both right ascension and declination.

\subsection{Data Reduction}

We followed the typical method for reducing near-IR camera data.  Dome flats were taken through each band in order to address the pixel-to-pixel sensitivity variations inherent in the detector.   A median dark frame of a corresponding exposure time was subtracted from the individual flats.  A median-combined flat field was then created for each filter.

The individual frames of the nine-point dither pattern for each source were inspected by eye for obvious problems and either accepted for subsequent processing or rejected.  A sky frame was produced for each source by applying a median filter to the dither pattern frames.  This sky frame was then subtracted from each dithered frame.  Then each sky-subtracted dithered frame was divided by the flat field (normalized to the mean).  We also explored an alternative technique in which the dithered frames were first divided by the normalized flat and then the sky subtraction was carried out.  This procedure produced results that were nearly identical to those obtained with the first reduction method. 

The individual frames were shifted to a common pixel position using the IRAF task IMSHIFT with a drizzle interpolant type and 0.5 pixel fractions.  The shifted frames were median-combined into a single image with the IMCOMBINE task set for the ccdclip rejection algorithm based on NIRIM's gain and readout noise parameters.

\subsection{Photometry}

Aperture photometry was carried out on the Herbig Ae/Be candidate stars using the APPHOT task in IRAF.  In all but one case, circular apertures were used to measure the analog-to-digital unit (ADU) counts from the source.  The aperture radius was increased until the ADU counts leveled off in order to ensure that the aperture fully captured the light from the source.  For the lone extended source, IRAS 19343$+$2926, the counts were measured using a polygon aperture.  Residual sky counts were measured in a concentric annulus and subtracted from the source ADU counts in the photometry annulus.  The sky-subtracted ADU count values were then converted into instrumental magnitudes by factoring in the exposure time and the usual flux-to-magnitude conversion.

Close binary candidate sources were put through the additional step of point-spread function (PSF) fitting in order to obtain more accurate photometry for the individual components.  The PSF fitting and photometry of the separated stars was carried out with the IRAF tasks PSF and ALLSTAR, respectively.  In practice, constructing model PSFs was complicated by the fact that the fields often contained few stars other than the target sources.  Therefore, the resulting instrumental magnitudes are likely limited by errors in the PSF model.

We attempted to correct for atmospheric extinction in the three bandpasses on each of the photometric nights.  This involved plotting the standard star instrumental magnitudes as a function of airmass and performing a linear fit to the data.  The slope of the fit line is then the airmass correction factor in units of mag airmass$^{-1}$.  The correction factor is applied to the standard stars to determine their airmass-corrected magnitudes.  In our case, the corrected standard star magnitudes were reduced to the zenith (airmass $=$ 1) instead of above the atmosphere values in order to conform to the technique used to obtain the published magnitudes \citep{hawarden01}.  The instrumental magnitudes of each of the program stars can also be corrected to their zenith value using the appropriate airmass value.

However, we were able to obtain airmass correction factors for only two of the photometric nights, 2003 December 6 and 2005 September 15.  The instrumental magnitude versus airmass plots from 2005 September 14 and 2005 September 16 either could not be fit with a line, or the trend was in the wrong direction, increasing airmass corresponded to a lower (brighter) magnitude.  This may be due to our sharing telescope time during the 2005 September run.  During partial nights, we were unable to observe standard stars over a wide range of airmass values.  It is also possible that occasional cirrus cloud cover went unnoticed.  Similar difficulties were encountered by \citet{ueta03} while carrying out photometric measurements with NIRIM.  The authors attributed the cause of this calibration problem to intrapixel sensitivity variations reported in NICMOS3 arrays \citep{lauer99}.  This effect can result in variations of up to 0.39 mag.  However, that explanation does not seem plausible for our data since the smallest PSF FWHM for all the runs was 0$\farcs$39.  This value covers about 2 pixels with NIRIM's plate scale on WIYN, and thus, the PSF is not undersampled.  Therefore, the exact cause of this problem remains unknown, but it resulted in airmass corrections not being applied to the 2005 September 14 and 16 data sets.

Nightly zero-point magnitudes for each bandpass were determined from the standard star instrumental magnitudes (either airmass-corrected or not depending on the observation date) and their published $JHK$ magnitudes.  The zero-point magnitude is defined as the magnitude that corresponds to a flux of one ADU count in 1 s.  These zero-point magnitudes were then used to compute the $JHK$$^{\prime}$ magnitudes of the Herbig Ae/Be candidate stars.  The resulting program star magnitudes are given in Table 3.  Sources with two entries are identified as close binary candidates with separate photometry for each star.  In addition, we present $H$-band 20$\arcsec$ $\times$ 20$\arcsec$ finding charts centered on the source in Figure 1.

Since most of our sources are relatively bright, the statistical errors for the magnitudes will be small, on the order of a few $\times$ 10$^{-3}$ mag.  However, the systematic errors from the airmass corrections and zero-point magnitude determinations will likely dominate the uncertainty.  In order to quantify the systematic error, we have carried out a comparison in which we derive the magnitude of a standard star based on its flux, and the flux and published magnitude of another standard star.  We then compute the difference between the derived magnitude of the first standard and its published value.  We adopted this value as the systematic error.  This procedure was carried out on all possible combinations of standard stars for each night and each band, and an average value was determined.  These nightly average uncertainty estimates, along with the photometric zeropoint magnitudes, for each bandpass are listed in Table 4.

\section{Results and Discussion}

\subsection{Binarity, Extended Structure, and Variability}

Our near-IR imaging has revealed five newly identified binary candidate systems among our program sources.  These sources are IRAS 05215$+$0225, 05295$-$0458, 05407$-$0501, 06013$-$1452, and 06531$-$0305.  We also resolve the components of IRAS 04278$+$2253, previously identified by \citet{white04} as a T Tauri binary system.  Separate photometry for each star in the candidate systems is given in Table 3.  The angular separation is defined as the distance between the coordinates found for each star by the PSF task in the $J$-band images.  If we exclude IRAS 04278$+$2253 and IRAS 19343$+$2926 (see below), these findings give us a binary fraction of approximately 15\%, less than that found in the previously-studied samples of \citet{leinert97} and \citet{baines06}.  However, our angular resolution ($\ga$0$\farcs$4) was not as good as what was achieved in these other observational studies (typically $\sim$0$\farcs$1).  We also applied a 2$\farcs$5 separation limit for two stars observed in relatively uncrowded fields to be considered a binary candidate.  According to \citet{artymowicz94}, the tidal truncation of a circumstellar disk due to a stellar companion sets an outer disk radius of approximately 1/3 the orbital separation distance.  We have computed disk outer radii based on this condition, our measured separations, and the distances reported in \citet{vieira03}, while assuming circular orbits viewed face-on.  These outer radii will be lower limits if the binary systems are inclined to our line of sight.  Most of the values fall within the range of outer radii for previously resolved circumstellar dust disks at near-IR wavelengths \citep[e.g.,][]{weinberger99, augereau01}.  The binary candidate system characteristics derived from this study are given in Table 5.  Adaptive optics (AO) or {\it Hubble Space Telescope} ({\it HST}) imaging of these binary candidates may be able to resolve any circumstellar/circumbinary dust associated with them, and permit a more detailed analysis of binary--disk interactions.  

We detect one extended source among our targets, IRAS 19343$+$2926 (M1-92, Minkowski's Footprint).  This source has been classified as a protoplanetary nebula (PPN) by \citet{bujarrabal98}, and therefore is likely not a Herbig Ae/Be star.  However, we still present its photometry in Table 3 and compare the results to previous measurements (see below).  We also report a secondary source approximately 3$\arcsec$ from the Herbig Be star HD 259431.  This source was not observed under photometric conditions and thus is not included in Table 3.  The brighter source was saturated and the image quality was poor, preventing us from giving a flux ratio or definitively concluding that the secondary source is in fact a stellar companion.

We have compared our photometric results to those obtained by 2MASS in order to search for near-IR variability.  Variations on the order of a few to several tenths of a magnitude are fairly common in our data set.  Unfortunately, making definite determinations about low-level variability ($\la$0.1 mag) in any given source is complicated, and often made impossible, by our relatively large uncertainties.  However, differences on the order of tenths of a magnitude likely indicate a real variation, making these sources excellent candidates for follow-up photometric monitoring.  The stars that show variability on this level when compared with the 2MASS data, excluding the binary candidates, are IRAS 05245$+$0022, F05272$-$0025, 05345$-$0139, 05355$-$0117, 05357$-$0650, 05417$+$0007, 06210$+$1432, 06245$-$1013, 06464$-$1644, 06491$+$0508, 06594$-$1113, 07020$-$1022, 07222$-2610$, 08277$-$3826, 19343$+$2926, and 20005$+$0535.  If we exclude the binary candidates and the PPN IRAS 19343$+$2926 from our statistics, the variable stars make up 54\% of the remainder of our sample.  We searched the 2MASS Survey, 6x, and Calibration Point Source Working Databases for multiepoch data on these sources as an additional check on photometric variability.  Multiepoch data were found for three sources, IRAS  05345$-$0139, 05417$+$0007, and 06210$+$1432.  Evidence for low-level variability is seen for each source in at least one band and is discussed for the individual stars in Section 3.2.

Most previous studies that address photometric variability in Herbig Ae/Be stars report optical variations in some of the sources \citep[e.g.,][]{herbig60, fink84, the94}.  In particular, \citet{fink84} report 18 variables, or suspected variables, in their catalog of 57 sources, or nearly 32\% of the sample.  Studies of near-IR variability in these objects remain relatively rare.  \citet{davies90} carried out near-simultaneous optical and near-IR photometry and measured variations in 15 of the 17 Herbig Ae/Be stars in their sample, including several found by \citet{fink84} to be not variable.  More recently, \citet{eiroa02} obtained simultaneous $UBVRI$ and $JHK$ photometry for a sample of PMS stars, and found that 9 out of 12 Herbig Ae/Be sources were variables.  The authors of these studies attribute the observed variability to photospheric activity, variable extinction, variable accretion, variations in scattered starlight due to disk structure, or combinations of these mechanisms.  Compared to these previous studies, our variable fraction is low.  However, their sample sizes were relatively small, and our fraction should be considered a lower limit since low-level variability may have gone undetected given our photometric uncertainties.  Clearly, further observational and modeling studies are needed to better understand the frequency and causes of photometric variability in the Herbig Ae/Be stars.

\subsection{Discussion of Individual Sources}

IRAS 03359$+$2932---Our $JHK$$^{\prime}$ magnitudes differ by 0.1 mag or less in all bands from the 2MASS photometry.  {\it HST} ACS coronagraphic imaging of this source did not reveal dust in optical scattered light \citep{doering07}.  A large ($\ga$ 800 AU) gas disk was detected by \citet{pietu03} with CO emission line observations.

IRAS 04278$+$2253---\citet{white04} have identified this source as a spatially resolved binary T Tauri system with spectral types of G8 and K7/M0.  Our imaging revealed a binary candidate with an angular separation of 1$\farcs$3 and $J$-band flux ratio of 1.8.  The combined $J$ magnitude (total $J$-band flux from both stars) is fainter than the 2MASS value by about 0.5 mag.  The $H$-band combined magnitude is fainter by nearly 0.6 mag.  Our position for this source is nearly the same as that reported by \citet{white04}.  It is relatively bright in the $J$-band and is found in an uncrowded field (see Figure 1).  Therefore, we conclude this is the same source that was observed by those authors, and we recommend it be considered a T Tauri binary.

IRAS 05044$-$0325---Our $H$ and $K$$^{\prime}$ magnitudes are in agreement with the 2MASS photometry.  The $J$ magnitudes differ by 0.19 mag, outside the reported uncertainties ($\sim$3.4$\sigma$ level).  This source is in a relatively uncrowded field, and it is not clear what physical mechanism would cause variation at this level only in the $J$ band.  Follow-up photometry is recommended.

IRAS 05136$-$0951---The NIRIM $J$ and $H$ magnitudes are nearly consistent with the 2MASS values.  The $K$$^{\prime}$ magnitude differs by about 0.14 mag, outside the reported uncertainties ($\sim$2.5$\sigma$ level).  One possible explanation for this discrepancy is that warm dust grains were created in a recent event.  However, this result should be confirmed.

IRAS 05215$+$0225---This source was identified as a binary candidate with an angular separation of 2$\farcs$4 and a $J$-band flux ratio of 25.8.  Our combined magnitudes are consistent with 2MASS in the $J$ and $H$ bands.  However, the $K$$^{\prime}$ total magnitude is 0.17 mag fainter than 2MASS.  \citet{hernandez05} have suggested that this source is surrounded by an optically thick inner disk.  Their finding was based on the object satisfying their criteria for selecting Herbig Ae/Be stars in their sample.  One of their criteria is a star's location on a $JHK$ color--color diagram; Herbig Ae/Be stars fall within a region that is well separated from the classical Be stars.  Our $JHK^{\prime}$ colors for the primary in this system differ slightly from the 2MASS colors that account for the contributions from both stars.  However, our colors still place the primary in the Herbig Ae/Be region on the color--color diagram.  Therefore, our photometric measurements, taken on their own, are also consistent with an inner disk associated with the primary when applying this \citet{hernandez05} criterion.  Additional observations will be required to determine whether the primary meets the authors'  emission line and spectral index criteria.

IRAS 05221$+$0141---Our $J$ magnitude is fainter than the 2MASS value by about 0.12 mag, while the $H$ and $K$$^{\prime}$ magnitudes agree with 2MASS.  Again, this behavior is not expected, and follow-up photometry is recommended.

IRAS 05245$+$0022---Our $JHK$$^{\prime}$ magnitudes differ from the 2MASS values by 0.35 -- 0.55 mag, with our magnitudes being consistently fainter.

IRAS F05272$-$0025---We report variations of about 0.1 -- 0.3 mag in the $J$, $H$, and $K$$^{\prime}$ bands.

IRAS 05275$+$1118---The $J$ and $H$ magnitudes are consistent within $\sim$1.5$\sigma$, and the $K^{\prime}$ magnitudes agree to well within $1\sigma$.

IRAS 05295$-$0458---We identify this source as a binary candidate with an angular separation of 2$\farcs$5 and a $J$-band flux ratio of 3.9.  Our combined $J$ and $K$$^{\prime}$ magnitudes are brighter than the 2MASS values by 0.91 and 0.45 mag, respectively.

IRAS 05345$-$0139---Our $J$ magnitude is about 0.4 mag fainter than the 2MASS value, and the $K$$^{\prime}$ magnitude is about 0.5 mag fainter.  Multiepoch 2MASS data consisting of two measurements separated by just over 689 days reveal deviations from constant magnitude values at the 2.1, 1.2, and 2.7$\sigma$ level in the three bands.

IRAS 05355$-$0117---We observe a variation of 0.17 mag between the NIRIM and 2MASS $H$ magnitudes, indicating a difference at the $\sim$2$\sigma$ level.

IRAS 05357$-$0650---We find a difference between the NIRIM and 2MASS magnitudes in the $J$ and $H$ bands of  0.3 and 0.8 mag, respectively.  In both cases, the NIRIM magnitudes are brighter.

IRAS 05407$-$0501---We identify this source as a binary candidate with an angular separation of 1$\farcs$1 and a $J$-band flux ratio of 5.2.  Our combined $J$ magnitude is in agreement with the 2MASS magnitude.  Our $K$$^{\prime}$ combined magnitude differs from 2MASS by almost 0.4 mag.

IRAS 05417$+$0007---There is a difference of nearly 0.2 mag in the $H$-band magnitudes obtained with NIRIM and 2MASS.  Multiepoch 2MASS data consisting of two measurements separated by just over 325 days reveals deviations from constant magnitude values at the 1.3, 1.1, and 2.2$\sigma$ level in the three bands.

IRAS 05598$-$1000---Our $J$-band magnitude for this source is in agreement with the 2MASS value.

IRAS 06013$-$1452---We have identified this source as a binary candidate with an angular separation of 1$\farcs$2 and a $J$-band flux ratio of 1.1.  Our combined $J$ and $K$$^{\prime}$ magnitudes are in agreement with the 2MASS photometry for this source.

IRAS 06040$+$2958---Our $H$-band magnitude is consistent with the 2MASS value for a source located almost 11$\arcsec$ from the J2000.0 coordinates as reported by Simbad.

IRAS 06111$-$0624---Our $J$, $H$, and $K$$^{\prime}$ magnitudes are consistent with the previously reported 2MASS values.

IRAS 06210$+$1432---This source is located in a crowded field, and our $J$ and $K$$^{\prime}$ magnitudes are fainter than the 2MASS values by about 0.3 and 0.6 mag, respectively.  Source misidentification or aperture contamination may have occurred.  Real variability is also a possibility.  Multiepoch 2MASS data consisting of two measurements separated by almost 174 days reveals deviations from constant magnitude values at the 3.4, 3.5, and 1.1$\sigma$ level in the three bands.  

IRAS 06245$-$1013---Our $J$ magnitude in this uncrowded field is about 1.2 mag fainter than the 2MASS value, and the $K$$^{\prime}$ magnitude is fainter by nearly 0.8 mag.  These values are well outside the uncertainties and imply variability.

IRAS 06464$-$1644---Our $J$ magnitude is consistent with 2MASS, while the $K$$^{\prime}$ magnitudes differ by almost 0.2 mag.  This may suggest a recent event in which grains were heated to temperatures over 1000 K.

IRAS 06475$-$0735---Our NIRIM $H$ magnitude agrees with the 2MASS value within the uncertainties.

IRAS 06491$+$0508---There is a 0.2 mag variation between the NIRIM and 2MASS $H$ magnitudes, with the NIRIM value being fainter.

IRAS 06523$-$2458---The NIRIM and 2MASS $H$ magnitudes agree within the uncertainties.

IRAS 06531$-$0305---We identify this source as a binary candidate with an angular separation of 1$\farcs$9 and a $J$-band flux ratio of 1.5.  Our combined $J$ magnitude is consistent with the 2MASS value.  The combined $H$ and $K$$^{\prime}$ magnitudes differ from the 2MASS photometry by 0.09 and 0.23 mag, respectively.

IRAS 06562$-$0337---The $J$ magnitude varies from the 2MASS value in this crowded field by 0.17 mag, a difference at the $\sim$2$\sigma$ level.  The $K$$^{\prime}$ magnitudes are consistent within the uncertainties.

IRAS 06594$-$1113---\citet{davies90} report an $H$-band variability of 0.07 mag, just above our uncertainty level.  However, the NIRIM $H$ magnitude is 0.27 mag fainter than the 2MASS value.  This may indicate that the amplitude of the variability is greater than the previously measured variations.

IRAS 07020$-$1022---This source has a reported $H$-band variability of 0.13 mag \citep{davies90}.  Our value differs from 2MASS by 0.26 mag, perhaps indicating a variability amplitude larger than previously measured variations for this source as well.

IRAS 07222$-$2610---There appears to be a 0.27 mag difference between the NIRIM and 2MASS $H$-band values, indicating variability.

IRAS 07303$-$2148---Our $H$-band magnitude is consistent with the corresponding 2MASS value for a source located nearly 12$\arcsec$ from this object's J2000.0 coordinates as reported by Simbad.

IRAS 08213$-$3857---Our NIRIM $H$-band magnitude is consistent with the 2MASS value.

IRAS 08277$-$3826---The NIRIM $H$-band magnitude differs from the 2MASS result by just over 0.2 mag, indicating possible variability.

IRAS 19343$+$2926---Otherwise known as M1-92, this extended source has been identified as a PPN by \citet{bujarrabal98} but was included in the \citet{vieira03} catalog.  While the $H$-band magnitudes are consistent between NIRIM and 2MASS, our $J$ magnitude is 0.5 mag brighter than the 2MASS value, and the $K$$^{\prime}$ magnitudes differ at the 1.8$\sigma$ level.

IRAS 20005$+$0535---Our $H$ magnitude differs from the 2MASS value by 0.6 mag for this uncrowded field, suggesting variability.

\section{Conclusions}

We have carried out a near-IR photometric imaging survey of Herbig Ae/Be candidate stars from the PDS.  The observations have provided us with $J$, $H$, and/or $K$$^{\prime}$ magnitudes for 35 sources.  Our subarcsecond imaging reveals five Herbig Ae/Be binary candidate systems and one previously-identified T Tauri binary among the 35 photometric sources.  We measure separations of $\sim$1$\arcsec$--2$\farcs$5 for these sources, and report separate photometry for each star.  Multiwavelength high-resolution imaging and separate SED construction for these systems is highly recommended.  A secondary source approximately 3$\arcsec$ from the Herbig Be star HD 259431 was also detected.  Unfortunately, this source was not observed during photometric conditions, and the primary source was saturated, preventing us from obtaining a flux ratio.  Also, the HD 259431 image does not permit us to definitively classify the secondary source as a stellar companion.  Clearly, follow-up high-resolution imaging of this source would be beneficial.  All sources appear point-like, with the exception of the extended source IRAS 19343$+$2926, previously classified as a PPN, and the above-mentioned HD 259431 if the secondary source proves not to be a companion, but is instead a feature associated with the known star.  Photometric variability is suggested for several sources when our magnitudes are compared to previous measurements.  However, conclusions about variability is complicated by our relatively large uncertainties.  We recommend follow-up photometric monitoring of these sources.  Finally, we provide $H$-band finding charts for future high-resolution imaging of these sources.

\acknowledgments

This work was supported by NASA grant NAG5-12595.  We thank the WIYN Telescope staff for their support of this work.  The WIYN Observatory is a joint facility of the University of Wisconsin-Madison, Indiana University, Yale University, and the National Optical Astronomy Observatories.  Visiting Astronomer, Kitt Peak National Observatory, National Optical Astronomy Observatories, which is operated by the Association of Universities for Research in Astronomy, Inc. (AURA) under cooperative agreement with the National Science Foundation.  This publication makes use of data products from the 2MASS, which is a joint project of the University of Massachusetts and the Infrared Processing and Analysis Center/California Institute of Technology, funded by the National Aeronautics and Space Administration and the National Science Foundation.  The DSSs were produced at the Space Telescope Science Institute under U.S. Government grant NAG W-2166. The images of these surveys are based on photographic data obtained using the Oschin Schmidt Telescope on Palomar Mountain and the UK Schmidt Telescope. The plates were processed into the present compressed digital form with the permission of these institutions.  This research has made use of the SIMBAD database, operated at CDS, Strasbourg, France.

{\it Facilities:} \facility{WIYN}.

\clearpage

\begin{figure}
\epsscale{.80}
\plotone{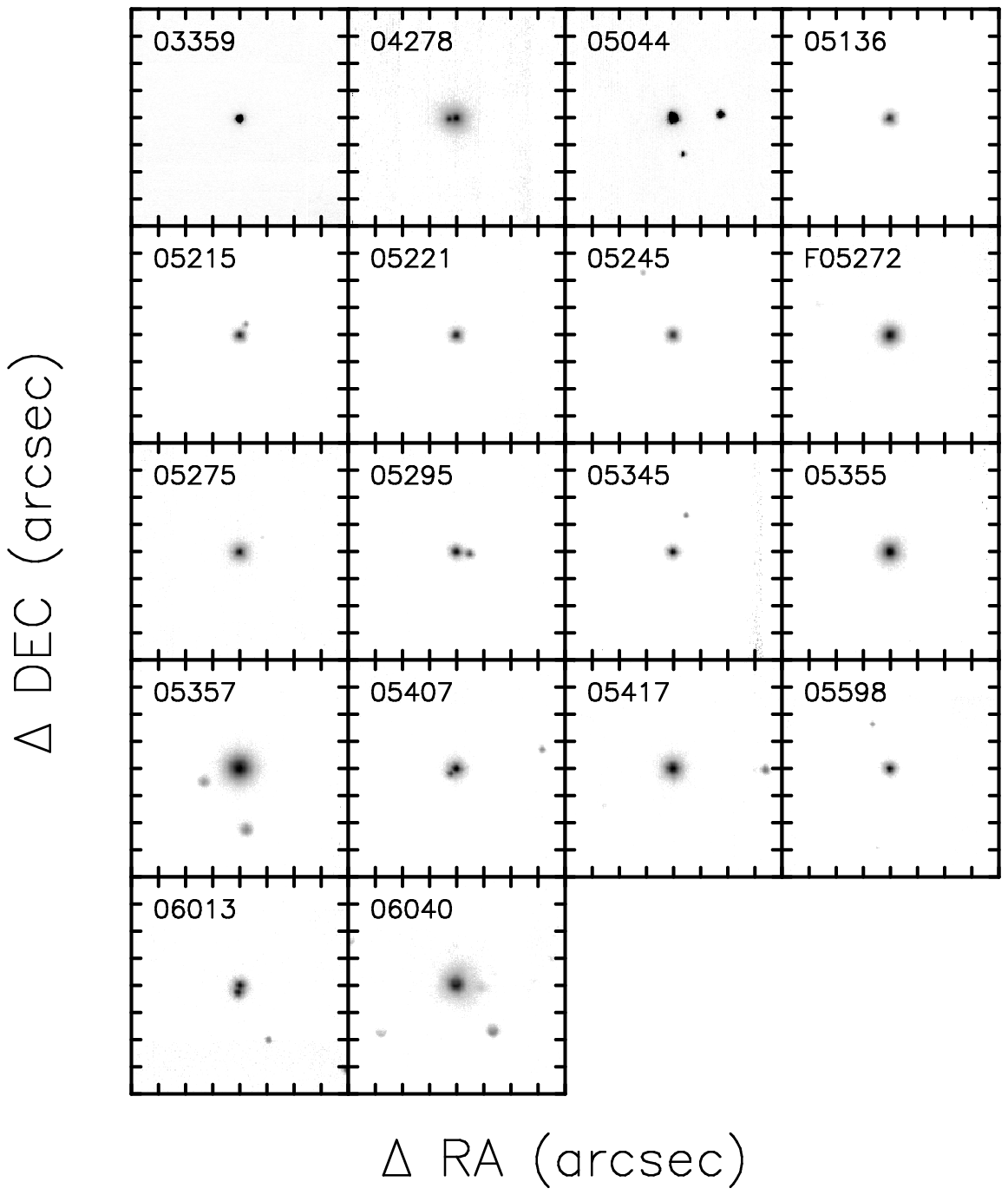}
\caption{$H$-band finding charts for the Herbig Ae/Be candidate sources.  Each frame is 20$\arcsec$ $\times$ 20$\arcsec$ and the source is centered in the field.  Tick marks are separated by 5$\arcsec$.  North is up and east is to the left.  The first five digits of the source's IRAS ID are given in the upper left corner of the frame.  IRAS 19343$+$2926 is clearly an extended source.  The extended structure seen in IRAS 06531$-$0305 is due to telescope drift during the observation.\label{fig1}}
\end{figure}

\clearpage

\addtocounter{figure}{-1}

\begin{figure}
\epsscale{.80}
\plotone{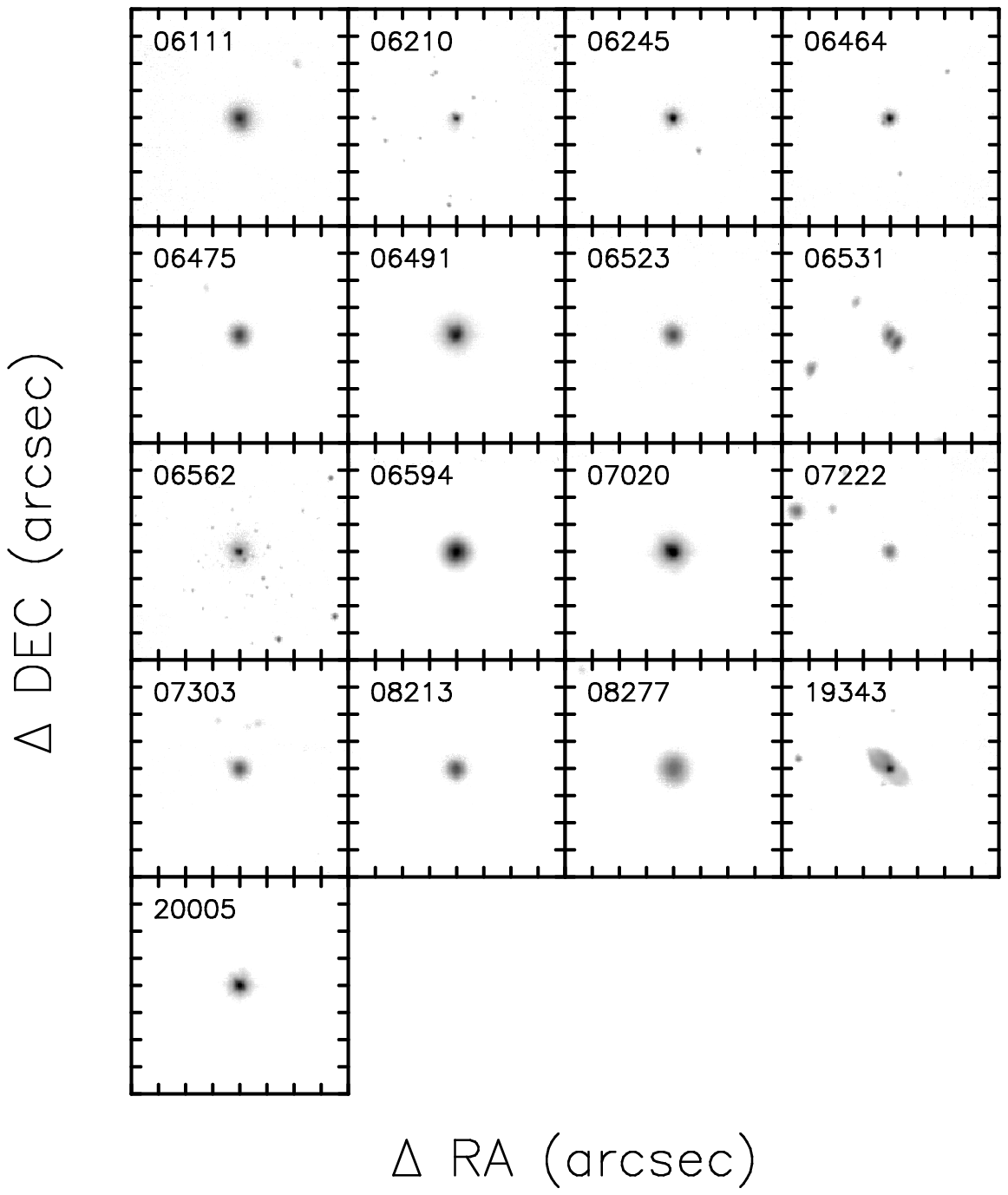}
\caption{{\it Continued}}
\end{figure}

\clearpage

\begin{deluxetable}{lccccc}
\tabletypesize{\scriptsize}
\tablecaption{Observation Log for NIRIM Imaging of Herbig Ae/Be Candidate Stars\label{tbl-1}}
\tablewidth{0pt}
\tablehead{ \colhead{Observing Run} &
\colhead{Date}   & \colhead{Average Seeing}  & \colhead{Conditions} \\  & & (arcsec) &}
\startdata
1               &  2003 Dec 05      &   0.49          &     Hazy    \\
                             &  2003 Dec 06      &   0.62         &     Photometric  \\
2               &  2004 Nov 26      &   1.07         &     Scattered clouds     \\
                             & 2004 Nov 27      &     \nodata          &     Cloudy     \\
3               &  2005 Sep 14    &    0.39          &      Photometric    \\
                             &  2005 Sep 15    &    0.50           &     Photometric     \\
                             &  2005 Sep 16    &    0.52          &     Photometric     \\
                             &  2005 Sep 18   &     0.58          &     Hazy     \\

\enddata

\end{deluxetable}

\clearpage

\begin{deluxetable}{lccccc}
\tabletypesize{\scriptsize}
\tablecaption{Near-IR Photometric Standard Stars$^{a}$\label{tbl-2}}
\tablewidth{0pt}
\tablehead{
\colhead{Star} & \colhead{R.A.} & \colhead{Decl.} & \colhead{$J$} & \colhead{$H$} &
\colhead{$K$}
}
\startdata
FS 109   & 03 13 24.16   & +18 49 38.4   &   11.432    &   10.968      &     10.807    \\
FS 111   & 03 41 08.55   & +33 09 35.5   &   10.640    &   10.378      &     10.289  \\
FS 112   & 03 47 40.70   &  -15 13 14.4   &   11.210    &   10.946      &     10.899     \\
FS 117   & 04 23 56.61   &  +26 36 38.0  &   11.465    &    10.515     &     10.078   \\
FS 120   & 06 14 01.44   &  +15 09 58.3  &   11.328    &   10.825       &     10.626     \\
FS 121   & 06 59 46.82   &   -04 54 33.2  &   11.984    &    11.423      &     11.315     \\
FS 124   & 08 54 12.07   &   -08 04 58.9  &   11.523    &    11.042      &     10.776     \\
FS 149   & 20 00 39.25   &  +29 58 40.0  &   10.106    &    10.083      &     10.086    \\
FS 155   & 23 49 47.82   &  +34 13 05.1  &   10.009    &      9.498      &       9.413     \\
\enddata
\tablecomments{Units of right ascension are hours, minutes, and seconds, and units of declination are degrees, arcminutes, and arcseconds (J2000.0).}
\tablenotetext{a}{Taken from Hawarden et al. (2001).}
\end{deluxetable}

\clearpage

\begin{deluxetable}{lccccccc}
\tabletypesize{\scriptsize}
\tablecaption{Near-IR Photometry of Program Stars\label{tbl-3}}
\tablewidth{0pt}
\tablehead{
\colhead{IRAS ID} & \colhead{Other Name} & \colhead{R.A.} & \colhead{Decl.} & \colhead{Date Observed} & \colhead{$J$} & \colhead{$H$} &
\colhead{$K$$^{\prime}$}
}
\startdata
03359$+$2932       & V1185 Tau                      &   03 39 00.6           & +29 41 46          & 2005 Sep 14  &   9.76    &   9.09      &     8.57      \\

04278$+$2253$^{a}$       & GSC 01829--00331     &   04 30 50.05         & +23 00 06.2      & 2005 Sep 14  &   9.78    &   7.90      &     \nodata       \\

      &           &                &                       &          &     10.39  &  9.18 & \nodata  \\

05044$-$0325        & NSV 1832                      &   05 06 55.5           &  -03 21 13          & 2005 Sep 14  &   10.08    &   9.13      &     8.48       \\

05136$-$0951        & HD 34282   &   05 16 00.4779    &  -09 48 35.421   & 2005 Sep 16 &     9.30    &      8.53      &       7.82    \\

05215$+$0225       & HD 287823 &   05 24 08.0493    & +02 27 46.886   & 2003 Dec 06 &   9.24    &   8.79       &     8.10     \\

&           &                &                       &          &     12.77  &  12.13 & 11.97   \\

05221$+$0141       & HD 287841 &   05 24 42.8015    &  +01 43 48.254  & 2005 Sep 16 &   9.82    &    9.18      &     8.52     \\

05245$+$0022$^{*}$       & HD 290409 &   05 27 05.475      &   +00 25 07.61    & 2005 Sep 16 &   9.86    &    9.50      &     9.18     \\

F05272$-$0025$^{*}$      & HD 290500 &   05 29 48.034      &  -00 23 43.16     & 2005 Sep 16  &   10.47    &    9.83      &     9.32  \\

05275$+$1118       & HD 244314 &   05 30 19.0           &  +11 20 20          & 2005 Sep 16  &   9.39    &      8.78      &       8.04   \\

05295$-$0458        & UY Ori          &   05 31 58.9           & -04 55 53             & 2003 Dec 06 &     10.82               &   \nodata                   &   8.99                \\

&           &                &                       &          &     12.29  &  \nodata & 11.46  \\

05345$-$0139$^{*}$        & HD 290770 &  05 37 02.449       & -01 37 21.36        & 2003 Dec 06 &  8.96                   &  \nodata                    &   7.59                 \\

05355$-$0117$^{*}$        & HD 290764 &  05 38 05.250       & -01 15 21.67        & 2003 Dec 06 &   \nodata                  &   8.03                   &   \nodata                  \\

05357$-$0650$^{*}$        & V1787 Ori                      &  05 38 09.6            & -06 49 15              & 2003 Dec 06 &  9.63                  &  8.16                     &   \nodata                  \\

05407$-$0501        & HD 38120   &  05 43 11.8928     & -04 59 49.897      & 2003 Dec 06 &      8.62             &  \nodata                      &  7.73                 \\

&           &                &                       &          &     10.42  &  \nodata & 9.46  \\

05417$+$0007$^{*}$       & HD 38238   &  05 44 18.7935     & +00 08 40.409     & 2003 Dec 06  & \nodata                  &    7.70                     &  \nodata            \\

05598$-$1000        & V791 Mon         & 06 02 14.9            & -10 00 59               & 2003 Dec 06 &    9.18               &   \nodata                      &   \nodata             \\

06013$-$1452        & AE Lep         & 06 03 37.0             & -14 53 03               & 2003 Dec 06 &    10.69                &  \nodata                      &   10.13             \\

&           &                &                       &          &     10.77  &  \nodata & 9.70  \\

06040$+$2958       & GSC 01876--00892                      & 06 07 16.1             & +29 58 01               & 2003 Dec 06 &    \nodata               &   9.44                      &   \nodata             \\

06111$-$0624        & GSC 04795--00492                      & 06 13 36.2             & -06 25 01                & 2003 Dec 06 &    10.34                &    9.85                    &   9.15              \\

06210$+$1432$^{*}$       & CPM 25        & 06 23 56.16          & +14 30 27.6            & 2003 Dec 06 & 11.92                   &   \nodata                      & 9.56              \\

06245$-$1013$^{*}$        & NSV 2968    & 06 26 53.8             & -10 15 34                 & 2003 Dec 06 & 11.26                   &   \nodata                    &     8.42          \\

06464$-$1644$^{*}$        & GSC 05990--00021                       & 06 48 41.8             & -16 48 06                & 2003 Dec 06 &    11.70                &   \nodata                    &   9.99              \\

06475$-$0735        & GSC 05379--00359                        & -6 49 58.8              & -07 38 52                 & 2003 Dec 06 &   \nodata                 &   10.33                    & \nodata           \\

06491$+$0508$^{*}$       & HD 50083    & 06 51 45.7531       & +05 05 03.865       & 2003 Dec 06 &    \nodata               &      6.74                &  \nodata        \\

06523$-$2458        & NSV 3275    & 06 54 26.8              & -25 02 12                & 2003 Dec 06 & \nodata                    &   10.54                  &  \nodata    \\                      

06531$-$0305        &  GSC 04805--01872                      & 06 55 39.3              & -03 09 45                & 2003 Dec 06 &   12.38                 &   11.33                 &   10.53      \\

&           &                &                       &          &     11.94  &  11.61  &   11.48  \\

06562$-$0337        & Ironclad Nebula      & 06 58 44.22            & -03 41 08.6             & 2003 Dec 06 &  11.34                  &   \nodata                &   9.95               \\

06594$-$1113$^{*}$        &  HD 52721   & 07 01 49.5104        & -11 18 03.328        & 2003 Dec 06 &  \nodata                   &  6.43              & \nodata          \\

07020$-$1022$^{*}$        &  HD 53367   & 07 04 25.5315        & -10 27 15.739        & 2003 Dec 06 &  \nodata                  &   6.48                  &  \nodata        \\

07222$-$2610$^{*}$        & GSC 06546--03156                       & 07 24 16.9               & -26 16 01                & 2003 Dec 06 &  \nodata                 &   11.76                   &  \nodata          \\

07303$-$2148        &  GSC 05991--00611                     & 07 32 27.2                & -21 55 27                & 2003 Dec 06 &   \nodata               &       10.82              &     \nodata        \\

08213$-$3857        &   GLMP 212                & 08 23 12.3                 & -39 07 03  & 2003 Dec 06 &  \nodata             &   8.68                   &   \nodata         \\

08277$-$3826$^{*}$        & HD 72106S & 08 29 34.8989          & -38 36 21.124       & 2003 Dec 06 &      \nodata          &    8.49                    &  \nodata         \\

19343$+$2926$^{b, *}$  & M1-92           & 19 36 18.82               & +29 32 51.4           & 2005 Sep 15 &  9.40                &  7.94                   &  6.32             \\

20005$+$0535$^{*}$       & HD 190073 & 20 03 02.5099         & +05 44 16.676       & 2003 Dec 06  & \nodata                 &      7.25                &  \nodata           \\
\enddata
\tablecomments{Units of right ascension are hours, minutes, and seconds, and units of declination are degrees, arcminutes, and arcseconds (J2000.0).  The stars marked with an asterisk are suspected variables.}
\tablenotetext{a}{This source has been identified as a T Tauri binary system by \citet{white04}.}
\tablenotetext{b}{This source has been identified as a PPN by \citet{bujarrabal98}.}
\end{deluxetable}

\clearpage

\begin{deluxetable}{lcccccc}
\tabletypesize{\scriptsize}
\tablecaption{Photometric Zeropoints and Estimated Systematic Uncertainties\label{tbl-2}}
\tablewidth{0pt}
\tablehead{ \colhead{Date} & \colhead{$Z_{mag}(J)$} & \colhead{$Z_{mag}(H)$} & \colhead{$Z_{mag}(K^{\prime})$} & \colhead{$\sigma_{J}$} & \colhead{$\sigma_{H}$} & \colhead{$\sigma_{K^{\prime}}$}   \\
       &    (mag)    &     (mag)   &   (mag) & (mag) & (mag) & (mag)  }
\startdata
2003 Dec 06    & 21.42 & 21.38 & 20.88 & 0.07   &  0.06   &   0.03        \\
2005 Sep 14   & 21.42 & 21.15 & 20.61 & 0.05   &  0.16   &   0.19        \\
2005 Sep 15   & 21.51 & 21.46 & 20.83 & 0.12   &  0.20   &   0.06         \\
2005 Sep 16   & 21.33 & 21.28 & 20.72 & 0.03   &  0.02   &   0.05         \\
\enddata
\end{deluxetable}

\clearpage

\begin{deluxetable}{lccccc}
\tabletypesize{\scriptsize}
\tablecaption{Close Binary Candidate Characteristics\label{tbl-2}}
\tablewidth{0pt}
\tablehead{
\colhead{IRAS ID} & \colhead{$d^{a}$} & \colhead{$\alpha$$^{b}$} & \colhead{$r^{c}$} & \colhead{$R_{out}^{d}$} & \colhead{$\Delta J$$^{e}$}  \\
                                         &   (pc)                    &   (arcsec)          &   (AU) & (AU)
}
\startdata
04278$+$2253   & 130--150   & 1.3   &   169--195    & 56--65  &  -0.61   \\

05215$+$0225   & 952   &  2.4  &   2285     & 762  & -3.53  \\

05295$-$0458   & 300--1600   &  2.5  &  750--4000 & 250--1333  &  -1.47      \\

05407$-$0501   & 422   &   1.1  &   464      & 155  & -1.8  \\

06013$-$1452   & 690--880   &  1.2  &   828--1056 & 276--352   & -0.08   \\

06531$-$0305   & 1300--1600   &  1.9  &    2470--3040  & 823--1013  & 0.44  \\

\enddata
\tablenotetext{a}{Distances taken from \citet{vieira03}.}
\tablenotetext{b}{Measured angular separation of binary candidate components.}
\tablenotetext{c}{Projected binary separation assuming the distances in Column 2.}
\tablenotetext{d}{Tidally truncated circumstellar disk outer radius.  The derivation of this value is described in the text.}
\tablenotetext{e}{The difference in the $J$-band magnitudes of the binary candidate components.}
\end{deluxetable}

\end{document}